INL/CON-08-14072
PREPRINT# Dissipative Particle Dynamics and Other Particle Methods for Multiphase Fluid Flow in Fractured and Porous Media

6<sup>th</sup> International Conference on CFD in Oil & Gas, Metallurgical and Process Industries

Paul Meakin
Zhijie Xu

June 2008

The INL is a
U.S. Department of Energy
National Laboratory
operated by
Battelle Energy Alliance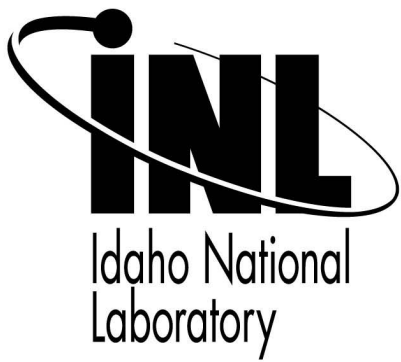

This is a preprint of a paper intended for publication in a journal or proceedings. Since changes may be made before publication, this preprint should not be cited or reproduced without permission of the author. This document was prepared as an account of work sponsored by an agency of the United States Government. Neither the United States Government nor any agency thereof, or any of their employees, makes any warranty, expressed or implied, or assumes any legal liability or responsibility for any third party's use, or the results of such use, of any information, apparatus, product or process disclosed in this report, or represents that its use by such third party would not infringe privately owned rights. The views expressed in this paper are not necessarily those of the United States Government or the sponsoring agency.

# DISSIPATIVE PARTICLE DYNAMICS AND OTHER PARTICLE METHODS FOR MULTIPHASE FLUID FLOW IN FRACTURED AND POROUS MEDIA

Paul MEAKIN[1,2,3]*, Zhijie XU[1]

[1] IDAHO NATIONAL LABORATORY Center for Advanced Modeling and Simulation
[2] INSTITUTE FOR ENERGY TECHNOLOGY Multiphase Flow Assurance Innovation Center, Kjeller
[3] PHYSICS OF GEOLOGICAL PROCESSES, University of Oslo.

- E-mail: Paul.Meakin@inl.gov

**ABSTRACT**

Particle methods are much less computationally efficient than grid based numerical solution of the Navier Stokes equation, and they have been used much less extensively, particularly for engineering applications. However, they have important advantages for some applications. These advantages include rigorous mass conservation, momentum conservation and isotropy. In addition, there is no need for explicit interface tracking/capturing. Code development effort is relatively low, and it is relatively simple to simulate flows with moving boundaries. In addition, it is often quite easy to include coupling of fluid flow with other physical phenomena such a phase separation. Here we describe the application of three particle methods: molecular dynamics, dissipative particle dynamics and smoothed particle hydrodynamics. While these methods were developed to simulate fluids and other materials on three quite different scales – the molecular, meso and continuum scales, they are very closely related from a computational point of view. The mesoscale (between the molecular and continuum scales) dissipative particle dynamics method can be used to simulate systems that are too large to simulate using molecular dynamics but small enough for thermal fluctuations to play an important role. Important examples include polymer solutions, gels, small particle suspensions and membranes. In these applications inter particle and intra molecular hydrodynamic interactions are automatically included

## INTRODUCTION

For most scientific and engineering applications the behaviour of fluids is simulated by numerically solving continuum equations such as the compressible or incompressible Navier Stokes equations. This is a highly developed approach that is based on familiar fluid properties such as the viscosity, surface tension and compressibility, or an equation of state for isothermal systems. However, fluids are composed of atoms and/or molecules, and it is generally accepted that the flow of fluids, including rheologically complex fluids could, *in principle*, be simulated by numerically integrating the classical equations of motion for the constituent particles. In additional to this molecular dynamics (MD) approach, a variety of particle methods have been developed for the purpose of simulating single- and multi-phase fluid flow. Familiar examples include dissipative particle dynamics (Hoogerbrugge and Koelman, 1992, Espanol and Warren, 1995) lattice gas models (Frisch *et al*., 1986) Monte Carlo methods (Bird, 1963), vortex particle methods (Cottett and Koumoutsakos, 2000), smoothed particle hydrodynamics (Gingold and Monaghan, 1977, Lucy, 1977) and the fluid-particle model (Espanol, 1998).

Except under extreme conditions, the flow of essentially all simple molecular fluids can be represented quite accurately by the Navier Stokes equation. This important universality arises because the intermolecular interactions conserve momentum, mass is conserved, simple fluids are isotropic and the behaviour of fluids is Galilean invariant because the particle-particle interactions depend only on *relative* positions. In addition, the Knudsen number must be small for continuum hydrodynamic behaviour. If the microdynamics of a particle model conforms to these basic conservation principles and symmetries, Navier Stokes fluid dynamics is virtually guaranteed on the continuum scale (if velocity dependent particle-particle interactions are included in the model these interactions must also be isotropic and depend only on velocity differences).

## PARTICLE MODELS

Particle models for fluid dynamics are based on numerical integration of the particle equation of motion, which has the form

$$m_i d\mathbf{v}_i / dt = \mathbf{f}_i = \sum_j \mathbf{f}_{ij} , \qquad (1)$$

where $m_i$ is the mass of particle $i$, $\mathbf{v}_i$ is its velocity, t is time and $\mathbf{f}_i$ is the force acting on particles $i$ due to its interactions with other particles and with solid boundaries and $\mathbf{f}_{ij}$ is the force acting on particle $i$ by particle $j$. In particle models, most of the computational effort is used to calculate the interaction forces. In MD simulations of specific chemical species the particle-particle interactions may be quite complex, and considerable effort is required to determine accurate interaction forces. However, in generic simulations of fluid dynamics the interactions should be as simple as possible, and in most cases simple pair-wise interactions, such as the Lennard-Jones interactions are used.

**Molecular Dynamics**

Molecular dynamics (MD) has been used to simulate both single-phase (Rapaport and Clementi, 1986, for example) and multiphase (Thompson and Robbins, 1989, for example) fluid dynamics. If realistic molecular interactions are used, the time step needed to obtain accurate results is on the order of $10^{-15}$ seconds. A very large scale state-of-the-art molecular dynamics simulation would consist of $O(10^{15})$ particle time steps for a relatively simple pair-wise interaction potential. This would allow a system of $10^9$ particles to be simulated for $O(10^{-9})$ seconds. It is well known that liquids within a few molecular diameters of solid surfaces have a more ordered structure than that of bulk water (Israelachvili and Pashley, 1983), and the characteristic size of the fluid domain must be much greater than the thickness of the ordered surface layer to obtain results that could be applied to much larger geometrically similar systems. The effect of fluid ordering near solid surfaces and slip would be negligible (except for extreme cases) in a simulation with $O(10^9)$ particles. Slip at solid-liquid interfaces is another challenge to the application of molecular dynamics simulations to flow in confined systems. Slip effects are negligible if $L >> L_s$, where $L_s$ is the slip length and $L$ is a length characteristic of the size of the confined system in which flow is taking place. For strongly wetting liquids, the slip length is quite small (O(1 nm)), but for partially wetting liquids the slip length may be much larger - tens of molecular diameters for large contact angles (Barrat and Bocquet, 1999). To investigate multiphase fluid flow, a strain of at least 10 is required and this implies that the strain rate would be $O(10^{10}\ sec^{-1}$.) This extremely high strain rate is similar to the strain rates encountered in nuclear explosions, and it is sufficient to cause significant changes in the structure of simple fluids such as water. If the number of particles were reduces to $O(10^6)$, the strain rate of $O(10^7\ sec^{-1})$ in a $(10^{15})$ particle time step MD simulation would not be sufficient to cause significant changes in the fluid structure, but wall effects (slip and ordering) could be significant. Providing wall effects are not too large, an $O(10^6)$ particle $O(10^{-6})$ second MD simulation could be used to simulate multiphase fluid flow in fractured and porous media with realistic capillary numbers, Bond numbers and Reynolds numbers. However, simulations on this scale are far from routine, and the effects of slip, ordering near solid surfaces and the strain rate are more severe for smaller scale simulations.

A "thermostat" is commonly used to maintain a constant temperature in non-equilibrium MD simulations. For example, the Anderson thermostat (Anderson, 1980) maintains a constant temperature by randomly selecting a small fraction of the particles during each time step in the simulation and re-setting their velocities to a new velocity selected randomly from the Maxwell-Boltzmann distribution for the required temperature. Other thermostats have been used in molecular dynamics simulations including the Nose Hoover thermostat (Martyna, *et al*., 1992) and the Berendsen thermostat (Berendsen *et al*., 1984).

Intermolecular interactions consist of a short range repulsive component and a (relatively) long range attractive interaction. The attractive interactions are responsible for the cohesive nature of liquids, and the repulsive, (excluded volume) interactions are largely responsible for the structure of the fluid (the way that molecules pack together). The time step used in MD simulations can be substantially increased by using a particle-particle interaction that is much softer than those that are typical of molecular systems. This leads to an increase in the compressibility of the model fluid, but providing that the Mach number remains smaller than $\approx 0.1$, the effect on fluid dynamics simulations is usually small. The softer interaction potentials should also reduce the molecular ordering near to solid surfaces, but we are not aware of systematic investigations of this effect.

One of the challenges encountered in the simulation of multiphase fluid flow is the complex dynamics of the fluid-fluid-solid contact line and the associated contact angle. Although the contact angle is sometimes assumed to be given by the simple equation

$$\cos(\theta_1) = (\Gamma_{s1} - \Gamma_{s2})/\Gamma_{12} \qquad (2)$$

obtained from a thermodynamic argument, the contact angle typically exhibits complex hysteretic behaviour. Here $\theta_1$ is the contact angle measured in fluid 1, $\Gamma_{s1}$ is the specific interfacial energy (energy per unit area) between the solid and fluid 1, $\Gamma_{s2}$ is the specific interfacial energy between the solid and fluid 2 and, and $\Gamma_{12}$ is the specific interfacial energy between fluid 1 and fluid 2. One of the advantages of MD simulations is that they can be used to investigate contact angle/contact line behaviour (wetting behaviour), and much of our knowledge and understanding of these phenomena come from MD dynamics simulations.

Dispersions of small particles or very large molecules have a wide range of practical applications and they have complex rheological behaviors. A system consisting of a few particles with diameters of a few hundred nm is still beyond the capabilities of molecular dynamics simulations. Brownian dynamics (BD) simulations are based on the idea that the particles in colloidal systems interact with each other via direct particle-particle interactions, drag forces due to the motion of the particles relative to the fluid and random (Brownian) forces resulting from momentum exchange with the molecular fluid. The BD equation of motion is

$$\mathbf{f}_i = \sum_{j \neq i} \mathbf{f}_{ij} + \gamma(\mathbf{v}_f - \mathbf{v}_i) + \boldsymbol{\xi}_i(t), \qquad (3)$$

where $\mathbf{f}_{ij}$ is the force acting on particle *i* due to its direct interaction with particle *j*, $\gamma(\mathbf{v}_f - \mathbf{v}_i)$ is the drag force acting on the particle due to its motion through the fluid ($\mathbf{v}_i$ is the particle velocity and $\mathbf{v}_f$ is the velocity of the fluid in the vicinity of the particle) and $\xi_i(t)$ is the random force acting on the particle because of thermal fluctuations. The random forces are related to the friction coefficient, $\gamma$, through the fluctuation-dissipation theorem (Kubo, 1966)

$$\langle \xi_i(t)\xi_i(t') \rangle = 6\gamma k_B T \delta(t-t'), \tag{4}$$

where $k_B$ is the Boltzmann constant and *T* is the absolute temperature.

Hydrodynamic interactions can be added to the standard BD equation of motion (Ermak and McCammon, 1978). However, this adds a substantial computational burden. The fluctuating and dissipative forces, related through the fluctuation-dissipation theorem, act as a thermostat, and Brownian dynamics simulations are very similar to thermostatted MD simulations.

**Dissipative Particle Dynamics**

Dissipative particle dynamics (Hoogerbrugge and Koelman, 1992; Espanol and Warren, 1995) can be thought of as a coarse-grained molecular dynamics model (Flekkoy and Coveney, 1999; Flekkoy *et al.*, 2000). Each particle represents a "cluster" of atoms or molecules, and, as a result of the internal degrees of freedom associated with each particle, the particle-particle interactions include random and dissipative contributions. In addition, the particle-particle interactions are relatively soft. The soft particle-particle interactions and the larger physical size of the particles are responsible for the high computational efficiency of dissipative particle dynamics (DPD) relative to MD, but there is little difference in practice between DPD and non-equilibrium (thermostatted) MD with soft particle-particle interactions.

The DPD equation of motion is

$$m_i d\mathbf{V}_i / dt = \mathbf{f}_i^C + \mathbf{f}_i^D + \mathbf{f}_i^R \tag{5}$$

In models for single phase fluids, the conservative force, $\mathbf{f}^C$, between particles has a simple purely repulsive form such as $\mathbf{f}_{ij}^C = -(1 - r_{ij}/r_0)\hat{\mathbf{r}}_{ij}$ for $r_{ij} > r_0$ and $\mathbf{f}_{ij}^C = 0$ for $r_{ij} = |\mathbf{r}_{ij}| < r_0$, where $\hat{\mathbf{r}}_{ij}$ is the unit vector pointing from particle *i* to particle *j* ($\mathbf{r}_{ij} = \mathbf{r}_j - \mathbf{r}_i / |\mathbf{r}_i - \mathbf{r}_j|$), where $\mathbf{r}_i$ is the position of particle *i* so that

$$\mathbf{f}_i^C = \sum_{j \neq i} \mathbf{f}_{ij}^C = -\sum_{j \neq i} S_{ij}(1 - r_{ij}/r_0)\hat{\mathbf{r}}_{ij}, \tag{6}$$

where $S_{ij}$ is the strength of the interaction between particle *i* and particle *j*. The dissipative particle-particle interactions are given by $\mathbf{f}_{ij}^D = -\gamma W^D(r_{ij})(\mathbf{r}_{ij} \cdot \mathbf{v}_{ij})\hat{\mathbf{r}}_{ij}$ so that

$$\begin{aligned}\mathbf{f}_i^D &= \sum_{j \neq i} \mathbf{f}_{ij}^D = \\ &-\sum_{j \neq i} \gamma W^D(r_{ij})(\mathbf{r}_{ij} \cdot \mathbf{v}_{ij})\hat{\mathbf{r}}_{ij},\end{aligned} \tag{7}$$

and the random forces are given by $\mathbf{f}_{ij}^R = \sigma W^R(r_{ij})\zeta \mathbf{r}_{ij}$, where $\zeta$ is a random variable with a zero mean and a unit variance.

The random and dissipative interactions are related through the fluctuation dissipation theorem (Espanol and Warren, 1995) which requires that $\gamma = \sigma^2 / 2k_B T$ and $W^D(r) = (W^R(r))^2$. The combination of dissipative and fluctuating forces, related by the fluctuation-dissipation theorem (Kubo, 1966) acts as a thermostat, which maintains the temperature of the system, measured through the average kinetic energy of the particles, at a temperature of T, providing that the time step used in the simulation is small enough. Consequently, DPD is essentially equivalent to non-equilibrium, thermostatted, MD with soft pair-wise particle-particle interactions, and the computational methods developed for MD simulations (Allen and Tildesley, 1987, Rapaport, 1996) can be applied to DPD. This idea can be taken one step further (Lowe, 1999) by integrating the equation of motion with only the conservative forces $\{\mathbf{f}_i^C\}$ over the time step, $\Delta t$, and then thermalizing the relative velocities of a fraction, $f_T$, of the particle pairs separated by a distance of $r_0$ or less by randomly selecting the relative velocities from a Maxwell Boltzmann distribution and multiplying by $\sqrt{2}$ to convert the Maxwell Boltzmann distribution for the velocity of a single particle at temperature T to a relative velocity distribution function and preserving the average velocity of the particles to conserve momentum.

In most particle models, the equations of motion are integrated using methods that are based on the velocity Verlet algorithm (Swope *et al.*, 1982). The dissipative and fluctuating forces complicate the task of developing accurate integration algorithms for DPD applications, and modified velocity Verlet algorithms (Groot and Warren, 1997, Pagonabarraga *et al.*, 1998, for example) have been investigated. If a velocity Verlet algorithm is used to integrate the stochastic (Langevin) equations of motion the kinetic temperature calculated from the particle velocities may deviate from the thermostat temperature, and this deviation can be used to assess the accuracy of the simulation.

The standard DPD model employs purely repulsive interactions, and the DPD fluid is a gas. However, if the temperature is too small at a particular particle mass and density the system will form a Kirkwood-Alder solid (Kirkwood 1939, Alder and Wainwright, 1962). This and other limitations to the degree of coarse graining that can be used in DPD simulations witrhout generating unacceptable errors have been investigated by Pivkin and Karniadakis (2006). They also point out that the thermodynamically consistent non-ideal DPD fluid model of Pagonaberraga and Frenkel (2000) with multibody interactions generates a DPD fluid with less pronounced local structure than the standard DPD model, and this allows a greater degree of coarse graining. The coarse-grained molecular dynamics DPD model of Flekkoy *et al.*, (Flekkoy and Coveney, 1999; Flekkoy *et al.*, 2000) does not undergo a Kirkwood-Alder transition, but a practical implementation of this model has not yet been developed. DPD simulations with purely repulsive interactions have been used extensively to simulate the behaviour of multiphase fluids in confined systems. A more realistic approach is to use a combination of short range repulsive and (relatively) long range atractive particle-particle interactions. This can be based on equation (6)

$$\mathbf{f}_i^C = -\sum_{j \neq i}(S_{ij}^r(1-r_{ij}/r_0^r) - S_{ij}^a(1-r_{ij)}/r_0^a))\hat{\mathbf{r}}_{ij}, \qquad (8)$$

where $S_{ij}^r$ and $r_0^r$ are the strength and range of the repulsive interactions and $S_{ij}^a$ and $r_0^a$ are the strength and range of the attractive interactions ($r_0^a > r_0^r$). An alternative is to use a pair-wise interaction function with the form

$$E(r) = S^r W(r, r_0^r) - S^a W(r, r_0^a), \qquad (9)$$

where $W(r, r_0)$ is a smooth bell-shaped function with a support scale of $r_0$, similar to the smoothing function used in smoothed particle hydrodynamics. The corresponding conservative forces are:

$$\mathbf{f}_i^C = \sum_{j \neq i}(S^r W'(r_{ij}, r_0^r) - S^a W'(r_{ij}, r_0^a))\hat{\mathbf{r}}_{ij}, \qquad (10)$$

where $W'(r, r_0) = dW(r, r_0)/dr$ (Liu *et al.*, 2006)

The van der Waals (vdW) equation of state

$$(P + a/v^2)(v - b) = k_B T \qquad (11)$$

is based on the idea that the effective volume available to the atoms or molecules is reduced due to the short range repulsive particle-particle interactions (*b* is the excluded volume), and the effective pressure is increased by the relatively long range attractive interactions (*a* is the attractive energy density). In equation (11), *P* is the pressure and *v* is the molecular volume. Figure 1 shows three-dimensional liquid droplets generated by DPD simulations with attractive and repulsive interactions.

The equation of state for a quite wide variety of molecular fluids can be represented quite well by a vdW equation. Consequently, it is not be surprising if DPD simulations with attractive and repulsive interactions can also be represented by the vdW equation of state. This is illustrated in figure 2. To obtain figure 2, the pressure was calculated using the virial theorem [Tsai, 1979, Allen and Tildesley, 1987]

$$P = P_k + \rho/3V \sum_{j>i}(\mathbf{r}_i - \mathbf{r}_j) \cdot \mathbf{f}_{ij}^c, \qquad (12)$$

where $P_k$ is the kinetic (ideal gas) pressure ($Pk = \rho k_B T$), where V is the volume.

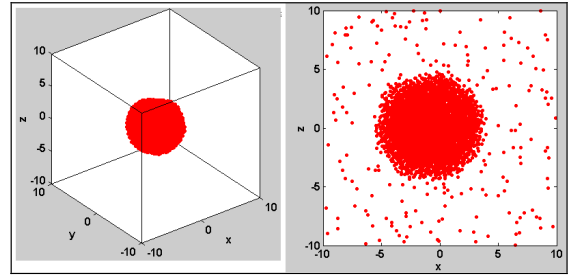

**Figure 1**: Three-dimensional simulation of a single component two phase fluid with interaction energies of $E(r) = 18.75(2W(r,0.8) - 1.05W(r,1.0)$ (left-hand-side) and $E(r) = 18.75(2W(r,0.8) - 1.0W(r,1.0)$ (right hand side). The energies are in units of $k_B T$ and $W$ is the most commonly used SPH smoothing function (see below). From Liu *et al.*, 2006.

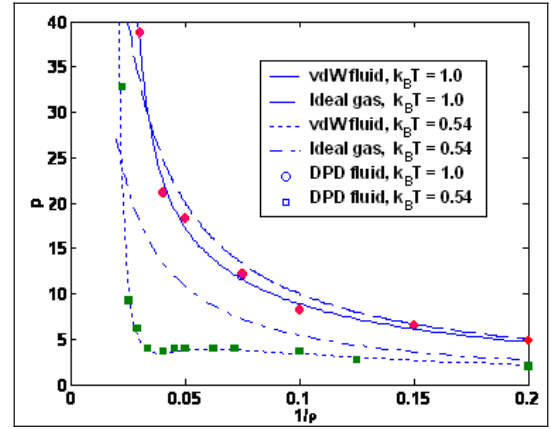

**Figure 2**: Equation of state for a vdW fluid is ($P = \rho k_B T/(1 - 0.016\rho) - 0.304\rho^2$, where $\rho$ is the fluid density). The particle-particle interaction energy is $E(r) = 18.75(2W(r,0.8) - 0.95(r,1.0)$.

$k_B T = 0.54$ for the single phase (supercritical) data set and $k_B T = 1$ for the two phase date set. From Liu *et al.*, 2006.

An alternative approach is to start with the vdW equation of state and derive the interaction forces from it (Pagonabarraga and Frenkel, 2001).

The particle-particle interaction approach can be used to simulated fluids confined to fracture apertures, pore volumes etc. The solid walls that confine the fluid(s) are represented by a thin layer of particles along the fluid-solid interface. A DPD simulation is first run in the entire computational domain, and the particles that are not within a preselected small distance, from the nominal solid-fluid interface are removed. The remaining particles are then immobilized, and they serve as the boundary particles. The interactions between the stationary wall particles and the fluid particles (as well as the wall roughness) control the wetting behaviour, and the repulsive part of the interactions between the fluid particles and the wall particles confines the fluid particles. Because of the soft nature of the particle-particle interactions, a few particles (those with high kinetic energies) can penetrate into the "solid" part of the computational domain. This is prevented by using bounce back boundary conditions at the fluid solid interface. Figure 3 illustrates the gravity driven motion of a drop of liquid in a straight narrow channel. In this simulation, and figures 4 and 5, the gravitational acceleration is 0.02. Figure 3 shows that the DPD model leads to realistic complex contact angle behaviour. However, this model cannot easily be used to simulate a specific system because the model parameters and the configuration of the wall particles cannot be theoretically related to either intermolecular interactions or the physical properties of solids and liquids. The fluid particle-particle interaction potential was $E(r) = 18.75(2W(r,0.73) - 1.05W(r,1.0))$, the temperature was $0.5k_BT$, and the dissipation strength was $\gamma = 4.5$. The strength of the fluid-solid interactions was twice that of the fluid-fluid interactions, and this results in wetting of the walls.

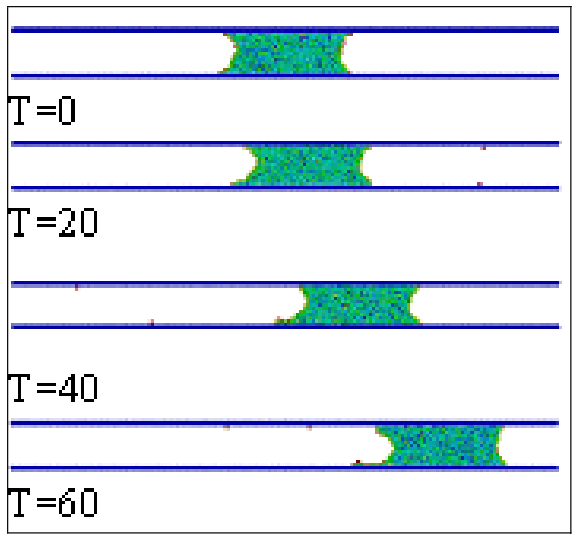

**Figure 3**: Two-dimensional DPD simulation of the gravity driven motion of a drop of liquid in a straight channel. Gravity acts from left to right.

Figure 4. shows a similar simulation performed using a channel with a complex geometry. The channel walls are self-affine fractals with a Hurst exponent of 0.7 (Mandelbrot, 1981, Meakin, 1998). This geometry was selected because a large body of experimental information indicates that the interfaces generated by the fracture of brittle materials such as rocks are self affine fractals (Schmittbuhl *et al.*, 1995, Daguier *et al.*, 1997). Comparison of figures 3 and 4 indicates that the increased roughness of the channel walls reduces the flow velocity, and a trailing film of liquid remains between the asperities. These effects are attributed to stronger pinning on the rougher walls.

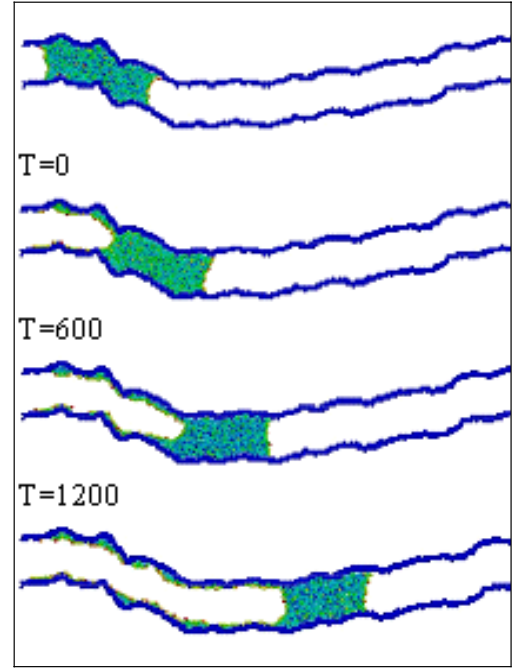

**Figure 4**: DPD Simulation of gravity driven flow through a geometrically complex channel. Gravity acts from left to right.

Figure 5 illustrated a two-dimensional DPD simulation of gravity driven flow across a fracture aperture. The entry of fluid into side branches and formation of liquid bridges across the side branches, similar to the behaviour seen in Figure 5 is observed in experiments and simulations carried out using other methods (Huang *et al.*, 2005). However, a detailed comparison of these two-dimensional DPD simulations with three-dimensional experiments and two-dimensional simulations with different geometries is not be justified.

It is difficult to compare simulations of multiphase fluid flow in fractured and porous media performed using DPD with simulations performed using Navier Stokes equation solvers without extensive "calibration" simulations to determine the viscosities and surface tension(s) of the DPD fluids. In addition, it is difficult to include the complex contact angle dynamics found in DPD simulation in continuum grid based models.

Figure 6. compares a DPD simulation with a finite volume solution of the Navier Stokes equation for the same geometry (a two-dimensional porous medium with a microfracture (Liu *et al.*, 2007)., In the DPD simulation $k_BT = 1$, the downward gravitational acceleration was 0.01 and the interaction energy was $E(r) = 18.75(2W(r,0.8) - W(r,1.0))$. The fluid-

fluid and fluid-solid interactions were equal. The volume of fluid model employed different advancing, receding and near-stationary contact angles, and the liquid properties were similar to those of water at standard temperature and pressure. The agreement between these simulations may be fortuitous, but it does suggest that DPD can be used to simulate multiphase fluid flow in fractured and porous media.

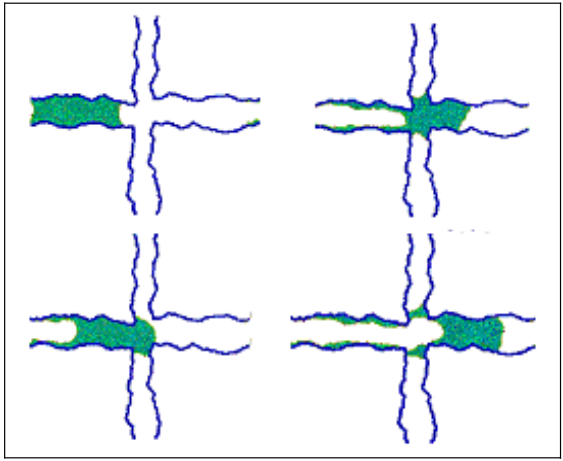

**Figure 5**: DPD Simulation of gravity driven flow through a "fracture aperture". Gravity acts from left to right.
.

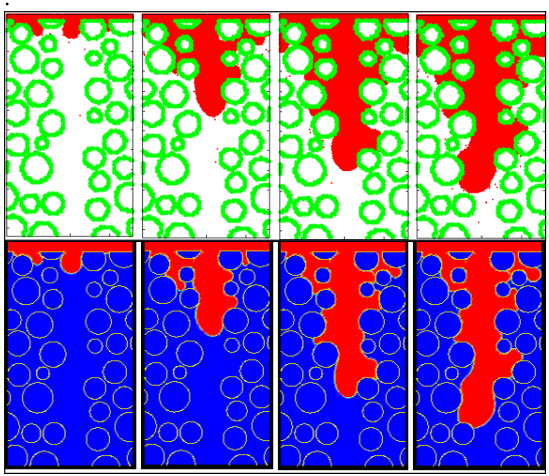

**Figure 6**: DPD (top) and finite volume CFD with volume-of-fluid interface tracking (bottom) simulation of gravity driven flow through a porous medium with a microfracture. Gravity acts from top to bottom. Based on Liu *et al*., 2007.

One of the advantages of DPD is that it is capable of simulating the flow of particle suspensions, polymer solutions, emulsions, gels and other complex fluids and soft condensed matter. An important application, which is outside of the usual scope of computational fluid dynamics is the effects of fluid flow on polymer conformation (Kong *et al*., 1997). Dissipative particle dynamics could also be used to simulate the fragmentation of polymers, other large molecules and colloidal aggregates due to thermal and hydrodynamic forces. To illustrate this, we show results from a simulation of the thermal decomposition of a "membrane" represented by a triangular network of unit masses connected by Hookian bonds. If the strain in a bond exceeds 0.1, the bond breaks irreversibly. In the simulation illustrated in figures 7 and 8, the bond breaking strain corresponds to an elastic energy of 1.0. Apart from the Hookian interactions, there were no DPD conservative forces, but the fluctuating and dissipative forces between bonded particles (nodes) were retained, and the range of these forces was 1.25 times the equilibrium bond length. Figure 7 illustrates a simulation performed at $k_BT = 0.2$. At this relatively low temperature, the removal of particles connected through only one bond is quite strongly favoured, and the membrane decomposes primarily at its edges leaving relatively large essentially undamaged regions.

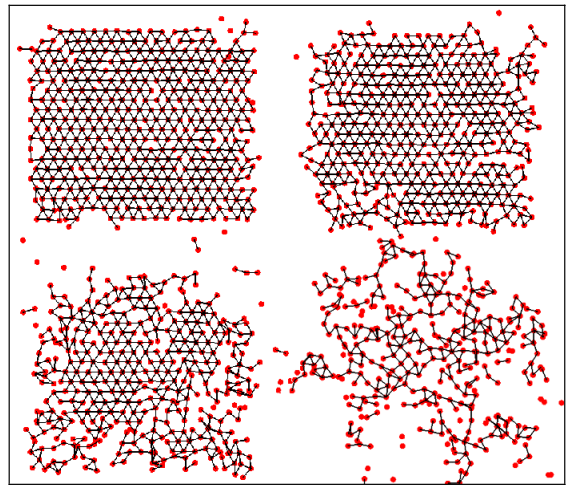

**Figure 7**: DPD Simulation of the thermal fragmentation of a two-dimensional membrane at a temperature of 0.2. Some of the detached nodes have moved out of the region shown (t = 26.0, 40.0, 57.0 and 77.0).

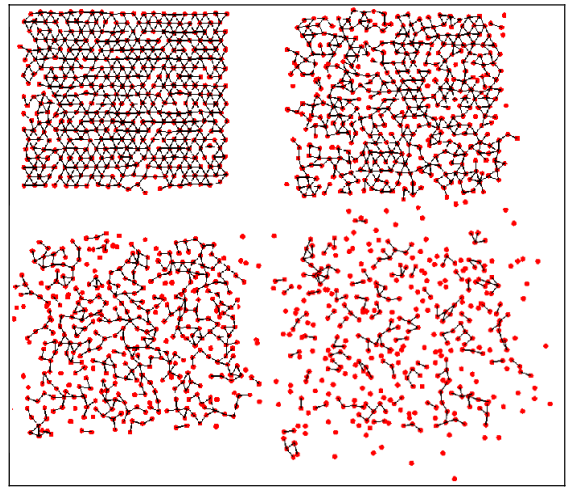

**Figure 8:** DPD Simulation of the thermal fragmentation of a two-dimensional membrane at a temperature of 0.5. Some of the detached nodes have moved out of the region shown (t = 1.2, 1.7, 3.1 and 4.5).

Figure 8 shows a similar simulation performed with $k_BT = 0.5$. At this higher temperature the ratios between the Boltzmann factors ($\exp(\delta E / k_BT)$ where $\delta E$ is the difference in the energy for two competing processes) are smaller, and the membrane decomposes much more rapidly and more uniformly. These simulations illustrate how DPD can be used to simulate processes that would be difficult to simulate using other models. The combination of fluctuating and dissipative interaction provides an accurate representation of the effects of thermal fluctuations.

**Smoothed Particle Hydrodynamics**

Smoothed particle hydrodynamics (SPH) was developed thirty years ago to simulate the fluid dynamics of astrophysical systems (Lucy, 1977; Gingold and Monaghan 1977). SPH is based on the idea that a continuous field, $A(\mathbf{r})$ can be represented by superimposing many smooth bell-shaped functions, $W(|\mathbf{r}-\mathbf{r}_i|)$, (the smoothing function or weighting function) centered on a set of points, $\{\mathbf{r}_i\}$. Similarly, the gradient of the field can be represented by the same superposition of the gradients of the smoothing function. A set of extensive properties, such as the particle mass, $m_i$, is associated with each particle, and these extensive quantities can be though of as being smoothed out or smeared by the smoothing function. The contribution of particle *i* to the fluid density field is given by $\rho^i(\mathbf{r}) = m_i W(|\mathbf{r}-\mathbf{r}_i|)$, and the smoothing function is normalized so that $\int W(\mathbf{r})d\mathbf{r} = 1$. Consequently, the density field is given by

$$\rho(\mathbf{r}) = \sum_i \rho^i(\mathbf{r}) = \sum_i m_i W(|\mathbf{r}-\mathbf{r}_i|). \qquad (13)$$

More generally, the intensive field $A(\mathbf{r})$ is given by

$$A(\mathbf{r}) = \sum_i A^i(\mathbf{r}) = \sum_i a_i W(|\mathbf{r}-\mathbf{r}_i|)$$
$$= \sum_i (m_i A_i / \rho_i) W(|\mathbf{r}-\mathbf{r}_i|), \qquad (14)$$

where $a_i$ is an extensive quantities carried by particle *i*. and the gradient of $A$ is given by

$$\nabla A(\mathbf{r}) = \sum_i (m_i A_i / \rho_i) \nabla W(|\mathbf{r}-\mathbf{r}_i|). \qquad (15)$$

In some applications it is advantageous to express the equations for continuous fields in terms of the particle number density (Tartakovsky and Meakin, 2006),

$$n_i = \sum_i W(|\mathbf{r}-\mathbf{r}_i|) \qquad (16)$$

and this leads to the expressions

$$A(\mathbf{r}) = \sum_i (A_i / n_i) W(|\mathbf{r}-\mathbf{r}_i|) \qquad (17)$$

$$\nabla A(\mathbf{r}) = \sum_i (A_i / n_i) \nabla W(|\mathbf{r}-\mathbf{r}_i|) \qquad (18)$$

The SPH equation for the flow of an inviscid fluid is based on the equation of motion

$$d\mathbf{v}/dt = -\nabla P / \rho, \qquad (19)$$

where $\mathbf{v}$ is the local fluid velocity and $\nabla P$ is the pressure gradient. First the pressure field is calculated from equation (11), and then the pressure gradient at each of the particles is obtained from the density field via the equation of state and the identity

$$\nabla P / \rho = \nabla(P/\rho) + (P/\rho^2)\nabla \rho \qquad (20)$$

The resulting equation of motion is

$$d\mathbf{v}_i/dt = -\sum_j m_j \left( \frac{P_i}{\rho_i^2} + \frac{P_j}{\rho_j^2} \right) \nabla W(|\mathbf{r}_i-\mathbf{r}_j|), \qquad (21)$$

where $\mathbf{v}_i$ is the velocity of particle *i*. This is one of many possible SPH formulations of the Euler equation for inviscid fluid flow. The point particles in the SPH method can be though of in terms of a moving disordered grid, and there are many ways of solving differential equations using SPH, just as there are many possible ways of formulating fluid flow equations using regular grids. In practice, equation (21) cannot be used to simulate the Euler equation because the particles in an SPH simulation move between regions with different velocities thus creating a viscosity due to momentum diffusion in the same way that viscosity is created in MD simulations (without including viscous forces in the equation of motion) and in simple fluids (Hoover, 1998).

If a body force, such as the effects of gravity acting on the fluid density is added, the equation of motion becomes

$$d\mathbf{v}_i/dt = \mathbf{f}_i/m_i - \sum_j m_j \left( \frac{P_i}{\rho_i^2} + \frac{P_j}{\rho_j^2} \right) \nabla W(|\mathbf{r}_i-\mathbf{r}_j|), \qquad (22)$$

where $\mathbf{f}_i$ is the body force acting on particle *i*. In many applications $\mathbf{f}_i = m_i \mathbf{g}$, where $\mathbf{g}$ is the gravitational acceleration.

Since the first applications of SPH were in the area of astrophysics, where viscous forces usually do not play a significant role, artificial viscosity was used to improve the numerical stability of SPH models, but little or no effort was made to develop SPH models for flows in which viscosity plays an important role, and it was almost 20 years until the effects of viscosity were included in SPH simulations (Takeda *et al*, 1994; Posch *et al.*, 1995). In SPH simulations, the effects of viscosity on fluid flow can be included by adding an SPH formulation of the viscous dissipation term in the

Navier Stokes equation, and the particle equation of motion can be expressed as

$$\frac{d\mathbf{v}_i}{dt} = \mathbf{g} - \sum_j m_j \left( \frac{P_i}{\rho_i^2} + \frac{P_j}{\rho_j^2} \right) \nabla W(|\mathbf{r}_i - \mathbf{r}_j|) + \sum_j \frac{m_j(\eta_i + \eta_j)(\mathbf{v}_i - \mathbf{v}_j)}{\rho_i \rho_j |\mathbf{r}_i - \mathbf{r}_j|^2} (\mathbf{r}_i - \mathbf{r}_j) \cdot \nabla_i W(|\mathbf{r}_i - \mathbf{r}_j|). \quad (23)$$

with one specific formulation of the viscosity term (Morris *et al.*, 1997; Zhu and Fox, 1997). Here $\mathbf{v}_i$ is the velocity of particle $i$ and $\eta_i$ is the fluid viscosity at particle $i$, (the viscosity can vary spatially in multiphase and/or multicomponent systems). The corresponding equation of motion based on the particle number density, $n$, is $m_i d\mathbf{v}_i / dt = \mathbf{f}_i$, where

$$\mathbf{f}_i = m_i \mathbf{g} - \sum_j \left( \frac{P_i}{n_i^2} + \frac{P_j}{n_j^2} \right) \nabla W(|\mathbf{r}_i - \mathbf{r}_j|) + \sum_j \frac{(\eta_i + \eta_j)(\mathbf{v}_i - \mathbf{v}_j)}{n_i n_j |\mathbf{r}_i - \mathbf{r}_j|^2} (\mathbf{r}_i - \mathbf{r}_j) \cdot \nabla_i W(|\mathbf{r}_i - \mathbf{r}_j|)_i. \quad (24)$$

While it is often convenient to use a Guassian form for the smoothing function in theoretical work, a variety of spline functions with a finite range, $h$, have been used in numerical work (Lucy, 1977; Gingold and Monaghan 1977; Zhu and Fox, 1999), and the smoothing function, $W(|\mathbf{r}|)$, in the above equations may be replaced by $W(|\mathbf{r}|, h)$ to emphasize this. The use of a finite range for the smoothing function is important in SPH simulations for the same reason that a finite interaction range is important in MD simulations - the order of the SPH algorithm is reduced from $O(n^2)$ to $O(n)$, where $n$ is the number of particles. The B-spline,

$$W(v,h) = \frac{\alpha_D}{h^D} \begin{cases} (1 - 3/2 v^2 + 3/4 v^3) & 0 \le v < 1 \\ 1/4 (2-v)^3 & 1 \le v < 2 \\ 0 & \text{otherwise} \end{cases} \quad (25)$$

is a widely used interpolation kernel (smoothing function), where $v = r_{ij}/h$, D is the spatial dimension, and $\alpha_D$ is a constant that assures the proper normalization of the smoothing function ($\alpha_D = 2/3, 10/7\pi, 1/\pi$ for D = 1, 2 and 3). This spline was used for the energy functions $W(r, r_0)$ in the DPD simulations discussed above.

One approach to simulating two phase single component fluids is to use a two phase equation of state, such as the vdW equation. However, this does not lead to stable liquid drops with smooth interfaces. This problem can be overcome by doubling the range of the smoothing function for the attractive forces corresponding to the long-range cohesive interactions in vdW fluids (Nugent and Posch, 2000), but this substantially increases the computational effort, particularly for three-dimensional simulations. An alternative approach is to add a combination of short range repulsive and (relatively) long range attractive particle-particle interactions, similar to those used to bring about phase separation in MD and DPD simulations (Tartakovsky and Meakin, 2005). This is illustrated in figure 9, which compares SPH and DPD simulations of an oscillating droplet.

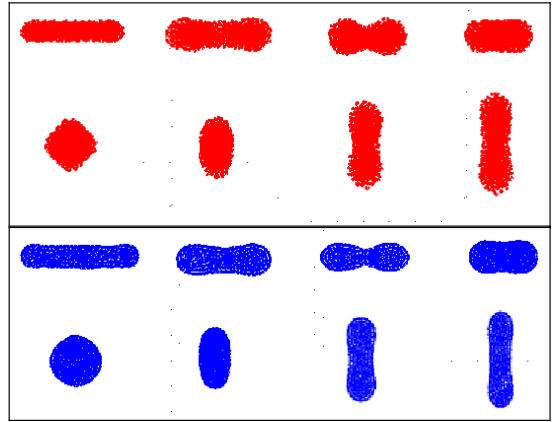

**Figure 9**: DPD (top) and SPH (bottom) simulations of an oscillating droplet. Adapted from *Liu et al*, 2006.

The DPD simulation was performed with $k_B T = 1.0$ and the particle-particle interaction potential was $E(r) = 18.75(2W(r, 0.8) - W(r, 1.0))$. The SPH simulation was based in a vdW equation of state. Again, the agreement is good, but it may be fortuitous because the fluid viscosities and the surface tensions were not determined.

Another approach, which we have recently investigated, is to base the SPH model on a phase field free energy functional (Ginzburg and Landau, 1950; Cahn and Hilliard, 1958). The force on each particle can be written as the gradient of the free energy,

$$\mathbf{f}_i = m_i d\mathbf{v}_i / dt = -\nabla_i E, \quad (26)$$

where $\nabla_i$ is the gradient with respect to the position of particle $i$. It is assumed that the total free energy is given by

$$E = \sum_{i=1}^{N} m_i \left( A_i(\rho_i, T_i) + \frac{k}{\rho_i} |\nabla \rho|_i^2 \right), \quad (27)$$

where $N$ is the number of particles and k is the magnitude of the gradient term in the Cahn-Hilliard Landau-Ginzburg free energy functional.

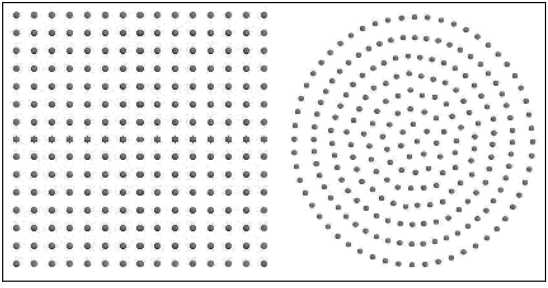

**Figure 10**: An arbitrary initial particle configuration and the final droplet obtained by minimizing the phase field/SPH free energy. A van der Waals equation of state was used.

The resulting equation of motion is

$$\frac{dv_i}{dt} = \sum_{j=1}^{N(j \neq i)} m_j \left\{ \left[ \left( \frac{P_i}{\rho_i^2} + \frac{P_j}{\rho_j^2} \right) + \left( \frac{k|\nabla\rho|_i^2}{\rho_j^2} + \frac{k|\nabla\rho|_i^2}{\rho_j^2} \right) \right] \frac{dW}{dr} \mathbf{e}_{ij} - \left[ \frac{2k(\nabla\rho)i}{\rho_i} - \frac{2k(\nabla\rho)j}{\rho_j} \right] \cdot \mathbf{S}_{ij} \right\} \quad (28)$$

where $\mathbf{e}_{ij}$ is the unit vector pointing from particle $j$ to particle $i$, and $\mathbf{S}_{ij}$ is given by

$$\mathbf{S}_{ij} = \nabla_i \left( \frac{dW}{dr_{ij}} e_{ij} \right) = \frac{d_2 W}{dr_{ij}^2} \mathbf{e}_{ij} \otimes \mathbf{e}_{ij} + \frac{1}{r_{ij}} \frac{dW}{dr_{ij}} (\mathbf{I} - \mathbf{e}_{ij} \otimes \mathbf{e}_{ij}) \quad (29)$$

where $\mathbf{I}$ is the unit matrix. The pressure, $P$, is obtained from the equation of state. Figure 10 illustrates this method.

## CONCLUSION

Although particle methods are much less computationally efficient than grid-based computational fluid dynamics, they have a number of important advantages for some applications such as the simulation of pore scale multiphase fluid flow in the subsurface. In the near future implementation of these models on high performance computing systems will allow simulations to be performed on a scale that permit constitutive relationships for use in continuum filed scale models to be obtained. Some particle models, such as dissipative particle dynamics can be used to simulate phenomena such as shear thinning, shear thickening, and viscosity reduction due to the fission of macromolecules that can represented only empirically in continuum models.

## REFERENCES

ALDER, B. J. and WAINRIGHT, T. E., (1962) "Phase transition in elastic disks", *Physical Review*, **127**, 359-361.

ALLEN, M. P. and TILDESLEY, D. J., (1987) "*Computer simulation of liquids*", Oxford Science Publications, Oxford, UK

ANDERSON, H. C., (1980) "Molecular dynamics simulations at constant pressure and/or temperature", *Journal of Chemical Physics*, **72**, 2384-2393.

BARRAT, J.-L. and BOCQUET, L., (1999) "Large slip effect at a non-wetting fluid-solid interface", *Physical Review Letters,* **82,** 4671-4674.

BERENDSEN, H. J. C., POSTMA, J. P. M., Van GUNSTEREN, W.F., DINOLA, A and HAAK, J.R., (1984). "Molecular-Dynamics with Coupling to an External Bath", *Journal of Chemical Physics,* **81,** 3684-3690.

BIRD, G. A. B., (1963), "Approach to translational equilibrium in a rigid sphere gas", *Physics of Fluids,* **6,** 1518-1519.

CAHN, J. W. and HILLIARD, J. E., (1958), "Free energy of a nonuniform system. I. Interfacial free energy", *Journal of Chemical Physics*, **28**, 258-267.

COTTET, G.-H. and KOUMOUTSAKOS, P., (2000), "Vortex methods: theory and applications", Cambridge University Press, Cambridge UK (2000).

DAGUIER, P., NGHIEM, B., BOUCHAUD, E. and CREUTZ, F. (1997), "Pinning and Depinning of Crack Fronts in Heterogeneous Materials", Physical Review Letters, **78**, 1062-1065.

ERMAK, D. L. and McCAMMON, J. A., (1978), "Brownian dynamics with hydrodynamic interactions", *Journal of Chemical Physics*, **69**, 1352-1360.

ESPANOL, P., (1998), "Fluid particle model", *Physical Review E*, **57**, 2930-2948.

ESPANOL, P. and WARREN, P., (1995), "Statistical dynamics of dissipative particle dynamics", Europhysics Letters, **30**, 191-196.

FAN, X., PHAN_THIEN, N, YONG, N. T., WU, X. and Xu, D., (2003), "Microchannel flow of a macromolecular suspension". *Physics of Fluids*, **15**, 11-21.

FLEKKOY, E. G., COVENEY, P. V. and De FABRITIS, G., (2000), "Foundations of dissipative particle dynamics" *Physical Review E*, **62**, 2140-2157.

FLEKKOY, E. G. and COVENEY, P. V., (1999), "From molecular dynamics to dissipative particle dynamics", *Physical Review Letters*, **62**, 1775-1778.

FRISCH, U., HASSLACHER, B. and POMEAU, Y. (1986) "Lattice-gas automata for the Navier Stokes equation", *Physical Review Letters,* **56,** 1505-1508.

GINGOLDS, R. A. and MONAGHAN, J. J. (1977) "Smoothed particle hydrodynamics: theory and application to non spherical stars", *Monthly Notices of the Royal Astronomical Society*, **181**, 775-389

GINZBURG, V. L. and LANDAU, L. D., (1950), "On the theory of superconductivity", *Zhurnal Eksperimentalnoy i Teoreticheskoy Fizicheskoi*, **20**, 1064-1082. (In Russian) English translation: Men of Physics: L.D. Landau (D. ter Haar, ed.), Pergamon, 1965, 138-167.

GROOT, R. D. and Warren, J., (1997), "Dissipative particle dynamics: Bridging the gap between atomistic and mesoscopic simulations", *Journal of Chemical Physics*, **81**, 4423-4435.

HOOGEBRUGGE, P. J. and KOELMAN, K. M. V. A. (1992) "Simulating microscopic hydrodynamic phenomena with dissipative particle dynamics", *Europhysics Letters*, **19**, 155-160.

HOOVER, W. G., (1998). "Isomorphism linking smooth particles and embedded atoms", *Physica A*, **260**, 244-254

HUANG, H., MEAKIN, P. LIU, M. B. and McCREERY, G. E., (2005) "Modeling of multiphase fluid motion in fracture intersections and fracture networks", *Geophysical Research Letters*, **32**, L19402.

ISRAELACHVILI, J. N. and PASHLEY, R. M. (1983) "Molecular layering of water at surfaces and origin of repulsive hydration forces", *Nature*, **306**, 249-250.

KIRKWOOD, J. G. (1939), "Molecular distribution in liquids", *Journal of Chemical Physics*, **7**: 919-925 (1939).

KONG, Y., MANKE. C. W. and MADDEN, W. G. (1997), "Effect of solvent quality on the conformation and relaxation of polymers via dissipative particle dynamics", *Journal of Chemical Physics*, **107**, 592-602.

KUBO, R., (1966), "The fluctuation-dissipation theorem", *Reports on Progress in Physics*, **29**, 255-282.

LIU, M. B., LIU, G. R. and LAM, K. Y., (2003) "Constructing smoothing functions in smoothed particle hydrodynamics with applications", *Journal of Computational Applied Mathematics,* **155**, 263-284.

LIU, M. B., MEAKIN, P. and HUANG, H. (2006), "Dissipative particle dynamics with attractive and repulsive particle-particle interactions", *Physics of fluids*, **18**, 017101.

LOWE, C. P., (1999), "An alternative approach to dissipative particle dynamics", *Europhysics Letters*, **47**, 145-151.

LUCY, L. B., (1977), "A numerical approach to the testing of the fission hypothesis", *The Astronomical Journal*, **82**, 1013-1024.

MANDELBROT, B. B., (1981), "The fractal geometry of Nature", Freeman, New York.

MARTYNA, G. J., RAPAPORT, D. C. and CLEMENTI, E., (1992), "Nose-Hoover chains: The canonical ensemble via continuous dynamics", *Journal of Chemical Physics*, **97**, 2635-2643.

MEAKIN, P. (1998), "*Fractals, scaling and growth far from equilibrium*", Cambridge University Press, Cambridge, UK.

MONAGHAN, J. J., (1992). "Smoothed particle hydrodynamics", *Annual Reviews of Astronomy and Astrophysics*, **30,** .543-574.

MORRIS, J. P., FOX, P. J. and ZHU, Y., (1997), " Modeling low Reynolds number incompressible flows using SPH", *Journal of Computational Physics,* **136**, 214-226.

NUGENT, S. and POSCH. H. A., (2000), "Liquid drops and surface tension with smoothed particle applied mechanics", *Physical Review E,* **62**, 4968-4975.

PAGONABARRA, I., HAGAN M. H. J. and FRENKEL, D., (1998), "Self-consistent dissipative particle dynamics model",. *Europhysics Letters*, **42**, 377-382.

PAGONABARRA, and FRENKEL, D., (2000) "Non-ideal DPD fluids. *Molecular Simulations*, **25**, 167-175.

PAGONABARRA, and FRENKEL, D., (2001) "Dissipative particle dynamics for interacting systems", *Journal of Chemical Physics*, **115**, 5015-5026.

PIVKIN, I. V. and KARNIADAKIS, G. E< (2006). "Coarse-graining limits in open and wall-bounded dissipative particle dynamics systems", Journal of Chemical Physics, 124, 1841012.

POSCH, H, A., HOOVER, W. G. and KUM, O., "Steady state shear flows via nonequilibrium molecular dynamics and smooth-particle applied mechanics", *Physical Review E*, **52**, 1711-1720.

RAPAPORT, D. C, (1996) "*The art of molecular dynamics simulations*", Cambridge University Press, New York.

RAPAPORT, D. C. and CLEMENTI, E., (1986), "Eddy formation in obstructed fluid flow: A molecular dynamics study", *Physical Review Letters*, **57,** 695-698.

SCHMITTBUHL, J., SCHMITT, F. and SCHOLZ, C., "Scale invariance of crack surfaces", (1995), *Journal of Geophysical Research-Solid Earth*, **100**. 5953-5973.

SWOPE , W. C., ANDERSON, H. C., BERENS, P. H and WILSON, K. R., (1982), "Computer simulation method for the calculation of equilibrium constants for the formation of physical clusters and molecules: application to small water clusters", *Journal of Chemical Physics*, **76**, 637-649.

TAKEDA, H., MIYAMA, SEKIYA, S. M., (1994), "Numerical simulation of viscous flow by smoothed particle hydrodynamics", *Progress in Theoretical Physics*, **92,** 939-960.

TARTAKOVSKY, A. M. and MEAKIN, P., (2005), "Modeling of surface tension and contact angles with smoothed particle hydrodynamics", *Physics Review E,* **72**, 026301.

TARTAKOVSKY, A. M. and MEAKIN, P., (2006), "Pore scale modeling of immiscible and miscible fluid flows using smoothed particle hydrodynamics", *Advances in Water Resources*, **29**, 1464-1478.

THOMPSON, P. A. and ROBINS, M. O., (1989) "Simulation of contact line motion – slip and the dynamic contact angle", *Physical Review Letters*, **63,** 766–769.

TSAI, D.H., (1979) "The virial theorem and stress calculation in molecular dynamics", Journal of Chemical Physics, 70, 1375-1382

ZHU, Y, and FOX, P. J., (1999), "Smoothed particle hydrodynamics model for diffusion through porous media", *Transport in Porous Media*, **43,** 441-471.